\documentclass{article}
\usepackage[english]{babel}
\usepackage[letterpaper,top=2cm,bottom=2cm,left=3cm,right=3cm,marginparwidth=1.75cm]{geometry}
\usepackage{amsmath,amssymb}
\usepackage{graphicx}
\usepackage[colorlinks=true, allcolors=cyan]{hyperref}
\usepackage{lipsum} 
\usepackage{authblk}
\usepackage{url}
\usepackage{tikz}
\usepackage{hyperref}
\usepackage{color}
\usepackage[normalem]{ulem} 
\usepackage{pbox} 
\usepackage{pgfplots}
\usepackage{makecell}
\usepackage{pbox}
\usepackage{tabulary}
\usepackage{array}
\usepackage{rotating}
\usepackage{enumitem,tikz}
\usepackage{multicol}

\pgfplotsset{compat=1.18} 
\graphicspath{{images/}}

\providecommand{\keywords}[1]
{
   	
  \textbf{{Keywords:}} #1
}

\title{A fast and automated approach for urban CFD simulations: integration with meteorological predictions and its application to drone flights}

\author[1,2,3,\thanks{corresponding author, email: \href{mailto:marcossuarez.vazquez@usc.es}{marcossuarez.vazquez@usc.es}}]{\underline{Marcos Suárez-Vázquez}}
\author[1,4]{Sylvana Varela Ballesta}
\author[2,3]{Alberto Otero-Cacho}
\author[2,3]{Alberto P. Muñuzuri}
\author[5]{Jorge Mira}

\affil[1]{Ventilatio Lab S.L., 15782 Santiago de Compostela, Spain}
\affil[2]{Galician Center for Mathematical Research and Technology (CITMAga), 15782 Santiago de Compostela, Spain}
\affil[3]{Group of Nonlinear Physics, Universidade de Santiago de Compostela, 15782 Santiago de Compostela, Spain}
\affil[4]{Universitat Rovira i Virgili, Departament d'Enginyeria Mecànica, 43007 Tarragona, Spain}
\affil[5]{Applied Physics Department and iMATUS, Universidade de Santiago de Compostela, 15782 Santiago de Compostela, Spain}

\begin{document}
\date{}
\maketitle
\begin{abstract}
    In past years, several studies have proposed new methods and applications for urban wind simulations. In this article, we present a fast and automatic methodology for reconstructing airflows within urban environments using LiDAR and cadastral data coupled with Computational Fluid Dynamics (CFD) simulations. Our approach integrates meteorological predictions with computational techniques to simulate the complex interactions between wind currents, buildings, vegetation, water zones and terrain morphology within urban environments. Accurate boundary conditions based on meteorological predictions are introduced into a coupled methodology that directly creates the terrain shape inside the simulation environment, simplifying the geometry creation process, which is one of the most prevalent problems in CFD urban simulations. The simulation results are confronted against ground-truth real data obtained from a meteorological station, showing strong agreement with the outcomes generated by the proposed CFD model, with a concordance correlation coefficient up to $\rho_c = 0.985$ for the wind direction and $\rho_c = 0.853$ for the wind speed. The results from these simulations are then used for validating a wind tunnel approach that mimics the interaction between a moving drone and the extracted wind currents, demonstrating a great improvement in computation times when compared to the most straightforward approach that consists in embedding the drone within the full urban landscape. This research contributes to the advancement of urban CFD modeling, and it has significant implications for various applications, providing valuable insights for urban development.
\end{abstract}


\section{Introduction}
\label{introduction}

Urban environments are characterized by complex aerodynamic challenges due to the intricate interactions between buildings, orography, vegetation, and atmospheric conditions \cite{liu2018influence, sandu2019impacts}. The rapid urbanization of densely populated regions across the globe has witnessed a significant growth in recent years \cite{toparlar2017review, angel2005dynamics, bocquier2005world}. Cities, as hubs of economic, social, and cultural activity, have grown exponentially. However, this surge in urbanization has also led to some challenges associated with large population concentrations. Urban climate is recognized as one of the main concerns of the next decade \cite{mirzaei2021cfd}, and some widely investigated topics include pollution dispersion in streets \cite{fernandez2023cfd, chu2005study}, energy demand of buildings \cite{allegrini2015coupled}, indoor/outdoor air quality \cite{munuzuri2022ventilation, yang2014cfd}, urban design optimization \cite{valger2019cfd} or pedestrian comfort \cite{blocken2012cfd, antoniou2019cfd}.

In the context of urban settings, Unmanned Aerial Vehicles (UAVs), commonly referred to as drones, have also seen a significant increase in use within large cities. The rapid expansion of the drone market \cite{de2018drone} has placed them at the forefront of various applications, including blood transportation \cite{amukele2017drone}, delivery of sensitive medical equipment \cite{pulver2016locating, haidari2016economic}, or animal tracking and detection \cite{koger2023quantifying, chabot2016computer}. Drones also play a crucial role in disaster management, assisting in flood and earthquake response efforts \cite{mishra2020drone, restas2015drone}, as well as in media broadcasting \cite{ayranci2016use}. Additionally, they have revolutionized package shipping and distribution \cite{benarbia2021literature}, significantly reducing costs, delivery times, and emissions compared to traditional alternatives \cite{rajabi2023drone}.

The modeling of wind patterns within urban environments has become an incredibly helpful asset to improve most of these applications. Computational Fluid Dynamics (CFD) simulations have emerged as a powerful tool for simulating airflow within urban areas, offering non-invasive insights into the generated airflow patterns, both for past experiments and future predictions. In the last years, a great number of building reconstruction tools like \cite{liu2017cfd, dhunny2018investigation, toparlar2018effect} have been proposed, even though they are not always adapted for CFD simulations. Some of them include flat terrains surrounding the buildings, which is a great disadvantage as demonstrated in \cite{brozovsky2021validation}, where the introduction of real terrain morphology is needed.

In \cite{paden2022towards, padjen2024automatic}, the authors propose a very useful method for the reconstruction of urban environments using LiDAR and cadastral data. However, this approach lacks the automation capabilities to be used for real-time predictions, as its use is restricted to the geometry generation, and lacks the implementation of automatic and complete simulations with accurate boundary conditions that correctly transmit low-resolution meteorological predictions into detailed simulations. Furthermore, challenges still persist in validating such models against real-world data, as highlighted in \cite{ju2021review}, who note that over 50\% of urban microclimate studies lack rigorous validation.

In recent years, some reference guidelines have been published for many aspects of urban simulations \cite{blocken2015computational}. Concepts such as the area of interest, which defines the optimal number of building blocks relative to a building of interest to achieve accurate simulations, and outer domain boundary sizes, are treated. Other aspects, such as the mesh type and its convergence study, the boundary conditions or the roughness parameter modeling have also been extensively studied, making urban CFD simulations more and more precise. However, this type of simulations are usually still very computationally expensive, needing large times for the generation and execution of precise simulations valid for real life applications as the ones previously mentioned.

In this article, we embark on a comprehensive study employing CFD simulations to analyze urban airflow dynamics, with a particular focus on leveraging meteorological predictions and observational data from real-life meteorological stations. The geometry generation is facilitated by the use of an algorithm that combines cadastral data and point cloud LiDAR technology, commonly used for reconstructing urban geometries \cite{alonso2010satellite}. Accurate boundary conditions are generated from the meteorological predictions and the influence of vegetation and water zones is incorporated into the airflow patterns of our simulations. This process is completely automated and can be applied to virtually any region in the world, with low execution times that allow its usage for applications where low response times are crucial. To account for an application of this methodology and further optimize computation times, the extracted wind field data from the urban simulation is used in an independent virtual wind tunnel simulation. This approach calculates aerodynamic parameters and load magnitudes over a drone geometry inside a small wind tunnel, which are then compared with the results from a direct, time-consuming simulation performed directly over a moving drone within the entire city reconstruction. The results from both simulations are compared, and the imporovement in computational cost is analyzed.

The primary aim of this research is to validate the accuracy of the proposed urban reconstructions and CFD simulations against observational data from meteorological stations, thereby enhancing confidence in computational predictions of urban airflow and assessing them in real-life scenarios, like a moving drone in a pre-planned path, where fast and accurate responses are needed under varying conditions. The predictions obtained by this method can be used for providing a better understanding of the airflow patterns that emerge in urban environments. This understanding can have a great utility for a wide range of applications, such as testing the viability of UAV flights under certain meteorological conditions. Other potential applications of this reconstruction methodology include optimizing ventilation in enclosed spaces by analyzing air direction and speed around buildings, as well as studying sound dispersion and acoustics in open areas, which are heavily influenced by obstacles and fluctuating air currents.

This paper is structured as follows. First, we present the geometry reconstruction methodology and describe the virtual wind tunnel developed for the drone. Next, we compare the simulated wind field with real-world data recorded at a meteorological station. Finally, we evaluate the wind tunnel approach using a generic drone geometry and compare its results with those obtained from a more computationally intensive simulation of the drone moving through an urban environment.

\section{Methodology}
\label{methodology}

\subsection{Numerical simulations using CFD techniques}
\label{theory}

Numerical simulations were carried out to analyze air circulation within the computational domain. The commercial software Simcenter STAR-CCM+ \cite{starccm} was used to design the domain, build the polyhedral meshes, solve the governing conservation equations (Navier-Stokes) using Finite Volume Methods (FVM) \cite{ferziger2019computational} and post-process the results. The equations for the conservation of mass and momentum \cite{kundu2024fluid} read as in Equations \eqref{eq:cons_mass} and \eqref{eq:cons_momentum},

\begin{equation}
	\label{eq:cons_mass}
	\frac{\partial}{\partial t} \int_V \rho dV + \oint_A \rho \mathbf{v} \cdot d\mathbf{a} = \int_V S_u dV,
\end{equation}
\begin{equation}
	\label{eq:cons_momentum}
	\frac{\partial}{\partial t} \int_V \rho \mathbf{v} dV + \oint_A \rho \mathbf{v} \otimes \mathbf{v} \cdot d\mathbf{a} = -\oint_A p\mathbf{I} \cdot d\mathbf{a} + \oint_A \mathbf{T} \cdot d\mathbf{a} + \int_V \mathbf{f}_b dV + \int_V \mathbf{s}_u dV,
\end{equation}
where $t$ is time, $V$ is volume, $\mathbf{a}$ is the area vector, $\rho$ is the density, $\mathbf{v}$ is the velocity, $p$ is pressure, $\mathbf{T}$ is the viscous stress tensor, $\mathbf{f}_b$ is the resultant of body forces and $S_u, s_u$ are user-specified source terms.

An incompressible solver with constant density was selected, along with a realizable k-$\varepsilon$ turbulent Reynolds-Averaged Navier-Stokes (RANS) model, considered suitable for numerical simulations of urban environments \cite{wang2006evaluation,parente2011comprehensive}. In addition, a two layer approach \cite{rodi1991experience} was used to gain flexibility of an all $-y^+$ wall treatment. SIMPLE algorithm was used for pressure-velocity coupling and a second-order upwind discretization scheme was chosen for the conservation of mass and momentum equations.

The transport equations for the kinetic energy $k$ and the turbulent dissipation rate $\varepsilon$ of the turbulence model are given as in Equations \eqref{eq:k} and \eqref{eq:epsilon},

\begin{equation}
	\label{eq:k}
	\frac{\partial}{\partial t} (\rho k) + \nabla \cdot (\rho k \mathbf{\overline{v}}) =
	\nabla \cdot \left[ \left( \mu + \frac{\mu_t}{\sigma_k} \right) \nabla k \right] 
	+ P_k - \rho \left( \varepsilon - \varepsilon_0 \right) + S_k,
\end{equation}

\begin{equation}
	\label{eq:epsilon}
	\frac{\partial}{\partial t} (\rho \varepsilon) + \nabla \cdot (\rho \varepsilon \mathbf{\overline{v}}) =
	\nabla \cdot \left[ \left( \mu + \frac{\mu_t}{\sigma_\varepsilon} \right) \nabla \varepsilon \right] 
	+ \frac{1}{T_\varepsilon} C_{\varepsilon1} P_\varepsilon 
	- C_{\varepsilon2} f_2 \rho \left( \frac{\varepsilon}{T_\varepsilon} - \frac{\varepsilon_0}{T_0} \right) + S_\varepsilon,
\end{equation}
where $\mathbf{\overline{v}}$ is the mean velocity; $\mu$ is the dynamic viscosity; $\sigma_k = 1.0$, $\sigma_{\varepsilon} = 1.2$, $C_{\varepsilon 1} = 1.44$, and $C_{\varepsilon 2} = 1.9$ are the model coefficients; $f_2$ is a damping function; $S_k$ and $S_{\varepsilon}$ are the user-specified source terms, and $P_k$ and $P_{\varepsilon}$ are the production terms defined by Equations \eqref{eq:prod1} and \eqref{eq:prod2},

\begin{equation}
	\label{eq:prod1}
	P_k = f_cG_k + G_b - \gamma_M,
\end{equation}
\begin{equation}
	\label{eq:prod2}
	P_{\varepsilon} = f_cS_k + C_{\varepsilon3}G_b,
\end{equation}
where $C_{\varepsilon3}=\tanh{(|v_b|/|u_b|)}$ is a model coefficient in which $u_b$ and $v_b$ are the velocity components parallel and perpendicular to the gravitational vector $\mathbf{g}$, $f_c$ is the curvature correction factor, $G_b$ and $G_k$ are buoyancy and turbulent production, $\gamma_M$ is the compressibility modification, $f_2 = k/\left(k+\sqrt{\nu \varepsilon}\right)$ is a damping function and $S_{k,\varepsilon}$ are the user-defined source terms.

\subsection{Geometry reconstruction}
\label{geom_prep}

In order to perform the reconstructed domain simulations, in which both the terrain orography and the presence of buildings play a crucial role, we first need to generate an appropriate geometry (manifold, watertight and completely closed), so that the CFD software can process it effectively. To achieve this, we developed an algorithm capable of reconstructing the geometries using information from two distinct datasets: LiDAR point cloud height data and cadastral information defining building footprints.

Fortunately, Spain provides well-documented and easily accessible LiDAR datasets that contain height information for every point (0.5-14 points/m$^2$), classified into various labels such as ground, vegetation, water, and buildings. This data, filtered using a Python script, provides a good notion of not only the height of a specific point, but also the buildings density in the desired area. Cadastral data is the other necessary dataset, as it helps defining the building footprints for the generation of precise geometries. For the case of Spain, it can be accessed and downloaded using a QGIS plug-in \cite{shurupov2023clasificador}, which provides a simple tool for downloading the information in an appropriate format.

The first step in our approach is the terrain reconstruction, from which the building geometries are then generated. A significant advantage of our all-in-one methodology is the direct integration between the Python code and the CFD simulations. By linking the code with Java macros inside Star-CCM+, we do not need to export the geometry before importing it into the software. This approach is particularly beneficial, as one of the main challenges in urban CFD simulations consists in preparing a closed and manifold geometry. Directly exporting the terrain would not only increase computational time but also introduce geometry errors that could require manual correction.

Instead of this, we generate a file that contains the height information of every terrain point within the domain. This file serves as a reference velocity for a Morphing motion inside the simulation, that deforms the domain to mimic the real terrain shape. As a result, real-life terrain features can be replicated by morphing the geometry inside the simulation software. Additionally, an .stl file with the shape of the ground is also generated, only for post-processing the results, as it helps visualizing velocity fields at a given height over the ground.

\begin{figure}[b!]
	\centering
	\includegraphics[width=1.0\textwidth]{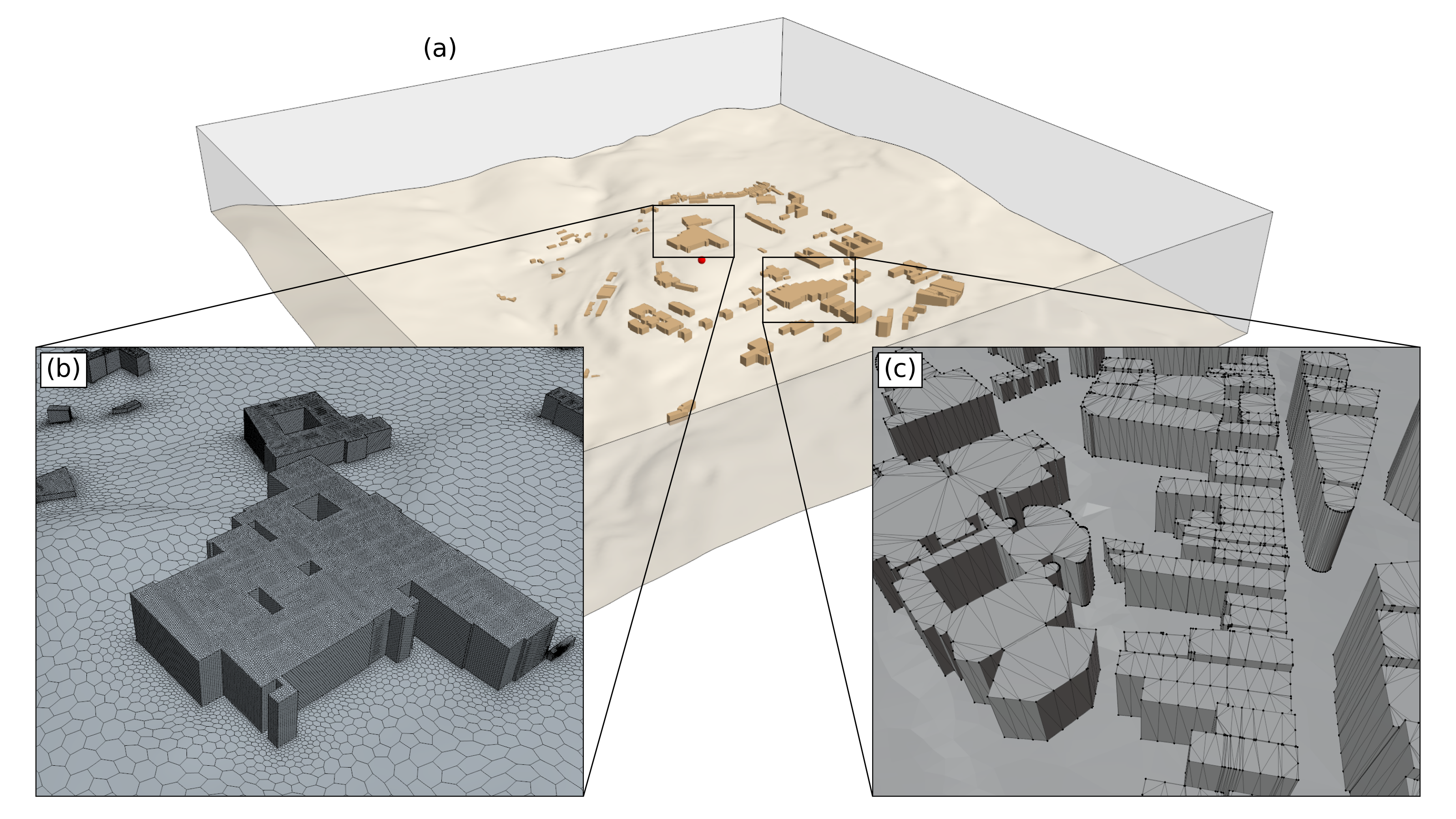}
	\caption{(a) Geometry used for the calibration process representing the Campus Sur in the University of Santiago de Compostela (Spain). The red dot represents the meteorological station from the official Galician weather service (MeteoGalicia), which we will use as a ground-truth reference value to assess the validity of our methodology. (b) Mesh used for the simulations. (c) Example of the resulting triangulation extracted from the Ear Clipping algorithm.}
	\label{fig:geometry}
\end{figure}

The 3D modeling of the buildings is constructed using a Level of Detail (LoD) of 1.2 according to the classification in \cite{biljecki2016improved}. This means that the buildings are reconstructed using the shape given by the cadastral dataset and raised to a uniform height equal to the mean height of the LiDAR points that fall into the building footprints. This level of detail corresponds to the approximated model in \cite{ricci2017local}, which maintains a good fidelity in terms of wind flows in comparison to a more detailed model. It should be stressed that higher LoD values than the one considered are way more complex to reconstruct, and more accurate data is needed. Figure \ref{fig:geometry} shows the geometry used for calibrating our methodology.

As we are working with an LoD 1.2 (i.e. buildings with flat roofs), in which the vertices of the top and bottom surfaces only differ in their $z$ coordinate, the greatest complexity for generating the geometry resides in triangulating the base of the building. For doing that, we chose to use the Ear Clipping algorithm \cite{eberly2008triangulation}, which is a method for triangulating a non-self-intersecting polygon. It works by iteratively identifying and removing ears---triangular sub-regions that can be clipped from the polygon without affecting its overall shape.

An ear is defined as a triangle formed by three consecutive vertices $(v_i, v_{i+1}, v_{i+2})$ of the polygon, where $v_{i+1}$ (the middle vertex) is convex, and the triangle does not contain any other vertices of the polygon inside it. The algorithm proceeds as follows:

\begin{enumerate}
	\item It starts with a simple polygon, represented as an ordered list of vertices. It identifies the convex vertices of the polygon, which are potential ear candidates.
	
	\item An ear is defined as a triangle formed by three consecutive vertices \( (v_i, v_{i+1}, v_{i+2}) \) of the polygon, where \( v_{i+1} \) (the middle vertex) is convex (its internal angle is less than 180°) and the triangle does not contain any other vertices of the polygon inside it. The algorithm scans through the polygon to detect all valid ears.
	
	\item A detected ear is removed from the polygon by eliminating its middle vertex \( v_{i+1} \). The remaining polygon is updated, and the process continues with the next available ear.
	
	\item Once an ear is clipped, the polygon shrinks. The new neighboring vertices are checked to see if they form new ears, and the process adapts to the updated shape of the polygon.
	
	\item Steps 2–4 are repeated until only one triangle remains in the polygon. The final result is a triangulated polygon, where all the interior regions are covered by non-overlapping triangles.
\end{enumerate}

This triangulation for the building bottom surface is then copied for the top face, and the lateral faces are easily calculated. This process is iterated for every building inside the domain, and the geometry is finally exported, as seen in Figure \ref{fig:geometry}(c). It is important to note that this process of generating the geometry is completely automated, as we only have to provide the coordinates of the region we want to reconstruct. This makes the entire process very fast, and once this information is provided, the whole process is automatic.

Even though the area of interest and outer domain sizes are completely customizable inside the configuration file, the dimensions are set in a way that meet the directional blockage ratios proposed in \cite{blocken2015computational}. This type of reference values are usually asigned to simple simulations with flat terrain and aligned buildings, but in real environments the analysis becomes a little harder. In our case, we select the tallest building (not the one situated in the highest part) and choose the minimum domain height (i.e., the distance between the highest point of the ground and the top surface) to reach a minimum blockage height ratio of 17\%, as recommended in \cite{blocken2015computational}. A similar rule is applied for the rest of horizontal ratios.

The number of needed building layers around the point we want to study, usually called area of interest, is a topic discussed in \cite{liu2018influence}. Again, as we are considering real-life morphologies, applying these reference values becomes harder, as we are sometimes not interested in a small area, but rather in a long path through various streets. For that reason, the area of interest can be modified freely inside the code, but some layers around the area of interest are always needed in order to get precise results in our paths.

Following this discussion, the reconstructed domain used for calibrating the model can be seen in Figure \ref{fig:geometry}. More examples of geometry reconstructions from different regions in Spain are provided in the Supplementary Material (Figures S.1, S.2 and S.3). In recent years, more countries have started performing LiDAR-based measurements of their territory and others have plans to do so, which means this method could be extended to reconstruct virtually any part of the world once the appropriate datasets are publicly available.

A mesh independency test was performed for this specific geometry. A polyhedral mesh with refinement near the buildings was applied. We tested five different base sizes in order to search for the best results with the lower number of elements for the mesh in Figure \ref{fig:geometry}(b). The most meaningful data for each simulation can be found in the Supplementary Material, where different magnitudes are compared in all meshes.

The Wall y+ values are high for the standard of typical CFD simulations, but still in range given the scales we are working with. Some authors suggest values lower than 1,000 for the majority of the cells, while others extend it to around 10,000 \cite{li2019cfd,aliabadi2018very,blocken2007cfd}. In our mesh, more than 50\% of cells have y+ values below 1,000, while up to 99.5\% of them are below 10,000. A more refined mesh near the buildings was considered in order to see if y+ values should have been optimized, but the wind speed and direction showed less than 1\% change, with 7 times the number of cells and the consequent computational cost. Therefore, we can conclude that high y+ values are acceptable when working with these spatial scales.

\subsection{Boundary conditions and modeling of vegetation}

To generate appropriate boundary conditions, we have to consider not only the wind speed provided by the meteorological service, but also the height at which it is provided. Intuitively, one can think of the progression of the wind speed as a boundary layer generated by the no-slip condition with the ground. To model this behavior, we will use a RANS model with a neutral Atmospheric Boundary Layer (ABL) logarithmic approximation \cite{parente2011improved}. A frequently used functional form for the velocity, turbulent kinetic energy and dissipation rate for neutral stratification conditions is:

\begin{equation}
	\label{eq:ABL_v}
	v(z) = \dfrac{u_*}{\kappa} \ln \left( \dfrac{z+z_0}{z_0} \right)
\end{equation}
\begin{equation}
	\label{eq:ABL_k}
	k = \frac{u^{*2}}{\sqrt{C_{\mu}}}
\end{equation}
\begin{equation}
	\label{eq:ABL_epsilon}
	\varepsilon = \frac{u^{*3}}{\kappa(z+z_0)}
\end{equation}
where $u_*$ is the friction velocity (whose value can be estimated by knowing a reference value for a given height), $\kappa=0.41$ is the von Karman constant, $z$ is the height above the ground level and $z_0$ is the roughness parameter, whose typical values depend on the type of surface considered \cite{wieringa1992updating}. The value of the roughness parameter controls the speed at which the velocity raises its value until reaching the condition imposed for $u_*$, so lower values for $z_0$ imply faster growing velocities.

\begin{figure}[b!]
	\centering
	\includegraphics[width=0.48\textwidth]{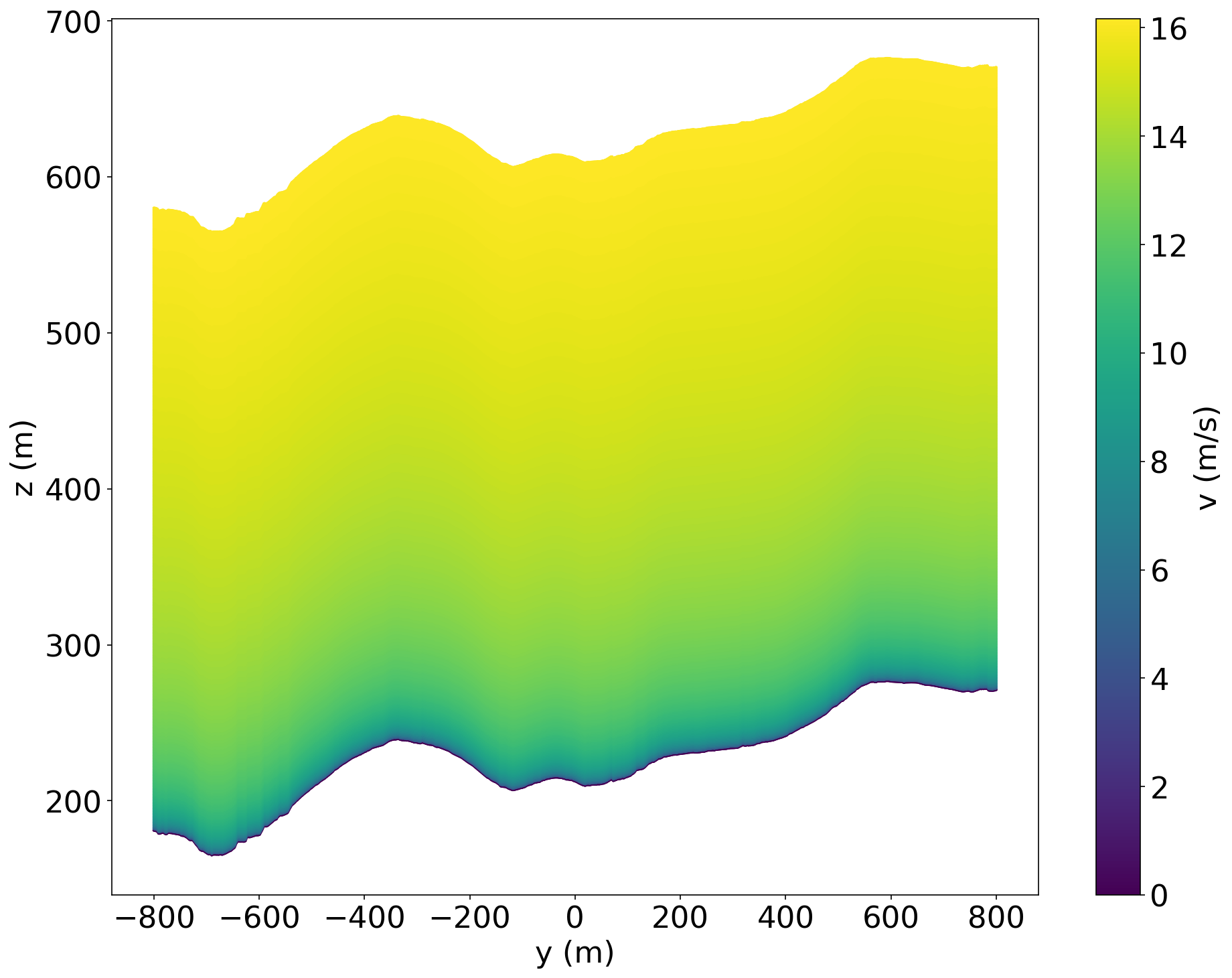}
	\includegraphics[width=0.48\textwidth]{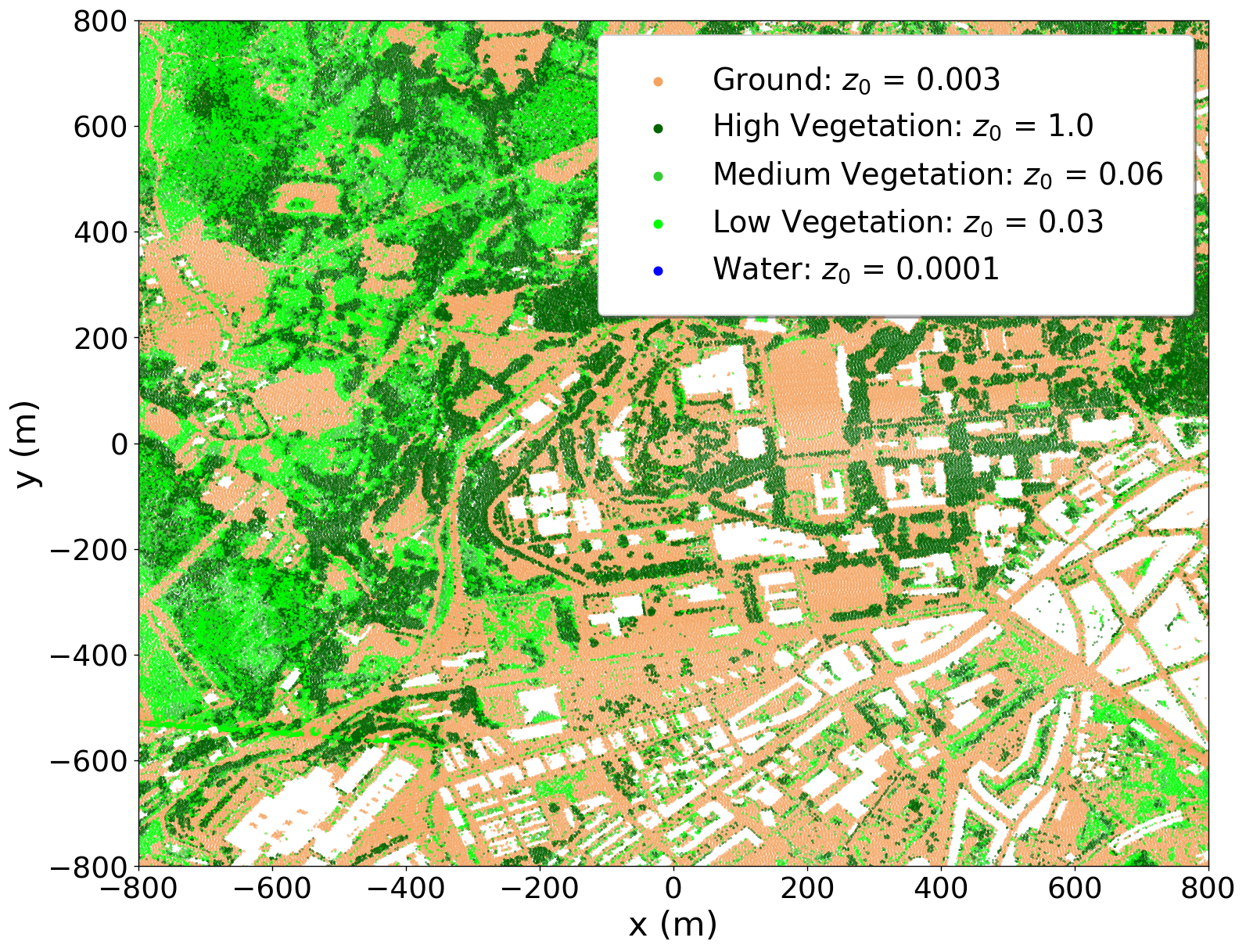}
	\caption{(a) Example of a boundary condition generated for the calibration geometry in the University of Santiago de Compostela. Note that the ABL equation for velocity is preserved at every point. (b) Vegetation zones inside the domain. The $z_0$ values are assigned at every point inside the simulation, mimicking the effect of vegetation on the velocity profile.}
	\label{fig:boundary}
\end{figure}

This concept of roughness parameter can be extended to model the presence of vegetation. As explained before, the LiDAR data is labeled with a distinction between low, medium and high vegetation. We can assign different $z_0$ values to every coordinate and apply them to the roughness function of the terrain. This will help us indirectly modeling the presence of different vegetation zones, and is particularly useful outside the area of interest, where the inlet boundary conditions have to completely develop into air currents. Different $z_0$ values produce distinct velocity profiles, which will affect the results once air reaches the region with buildings.

In this regard, some work around the modeling of vegetation and its effects in the results of urban CFD simulations has been done \cite{fu2024should}. However, directly modeling this vegetation usually requires high-precision data and is very difficult to automate. For this reason, we opted for an alternative approach, using a variable roughness wall parameter like in \cite{paden2022towards,di2022new}.

As said before, our code also takes the meteorological predictions at 10 m height from three different sources: the official Galician weather services (MeteoGalicia), the official Spanish weather service (AEMET) and OpenMeteo models. Using these predictions, it generates appropriate boundary conditions following the logarithmic approximation from Equation \eqref{eq:ABL_v}. This is not a trivial task, as we are not considering flat terrains near the boundaries. We opted for generating the boundary conditions that preserve Equation \eqref{eq:ABL_v} at every point.

Figure \ref{fig:boundary}(a) shows an example of one of the boundary conditions generated for the geometry in Figure \ref{fig:geometry}, along with an example of the roughness wall parameters used for the different labeled regions in Figure \ref{fig:boundary}(b).

\subsection{Wind tunnel for drone path}
\label{section:meth_wind_tunnel}

The most straightforward approach for testing a moving drone's behavior within a wind field derived from an urban CFD simulation is to embed the drone geometry directly into the simulation domain and simulate its motion through the flow. However, this method is highly computationally inefficient. It demands fine mesh resolution along the drone's entire path, greatly increasing the number of cells and further amplifying the already substantial computational cost of transient simulations.

\begin{figure}[b!]
	\centering
	\includegraphics[width=\textwidth]{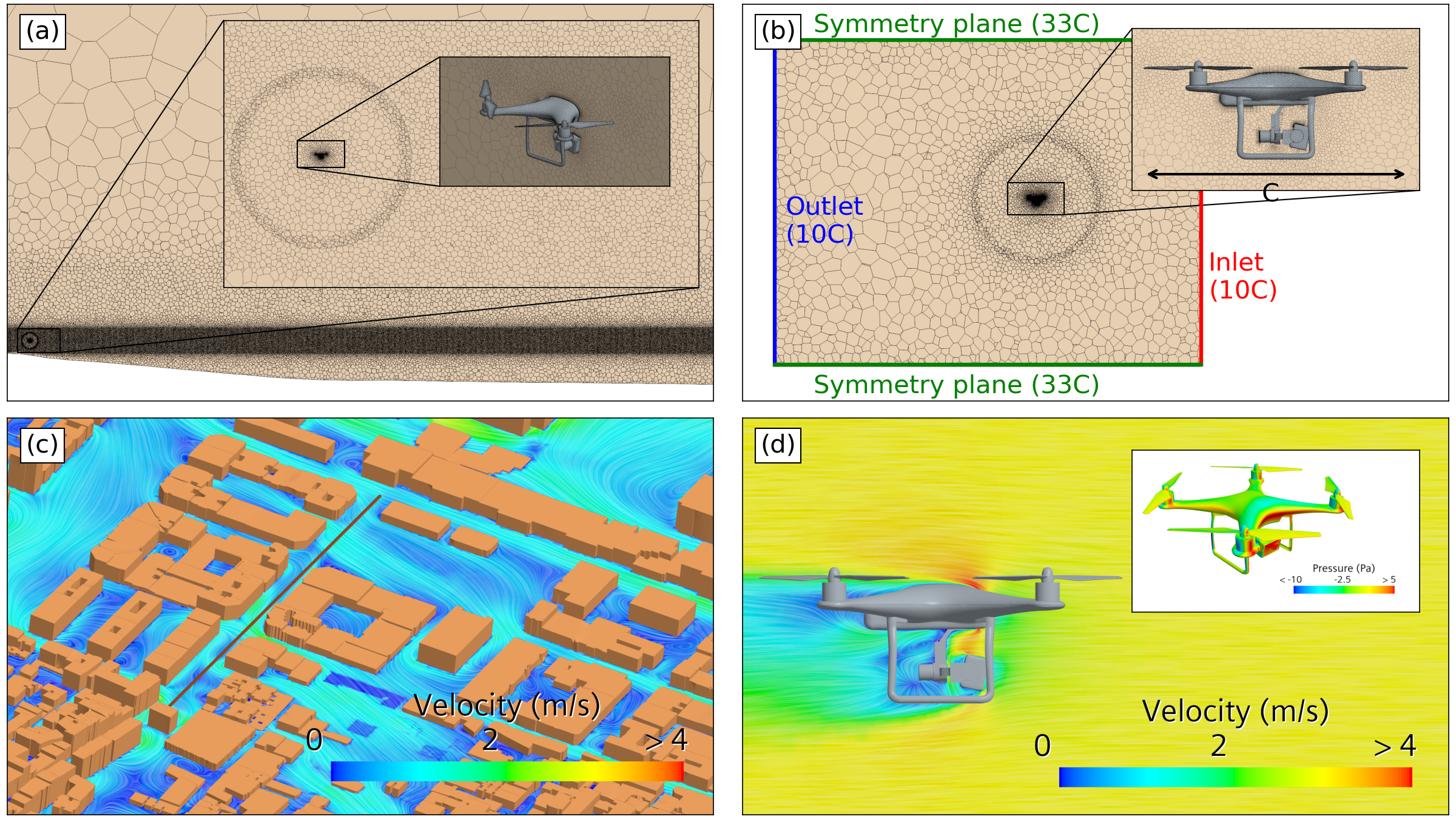}
	\caption{Overview of the simulation process. We start by selecting an appropriate mesh for both (a) the case of the complete domain over the city reconstruction and (b) the wind tunnel, with the same Base Size for the spherical region around the drone, extracted from a mesh independency test. The dimensions of the tunnel are described relative to the drone's diagonal wingspan $C = 0.4$ m, using the reference values from \cite{prabhakararao2014cfd, hassan2014numerical}. Then, (c) the whole city domain is initialized by running a  steady simulation, and the wind data is extracted following a given trajectory (brown line). A transient simulation with a moving drone in the whole domain is performed and compared with (d) the wind tunnel approach with lower computational times.}
	\label{fig:mesh_wind_tunnel}
\end{figure}

To address this limitation, we propose an alternative approach that decouples the drone simulation from the full urban CFD model. Specifically, we use the wind velocity field from a preliminary steady-state CFD simulation of the urban environment as input for a separate, transient simulation conducted in a smaller wind tunnel domain.

A key challenge of this approach is the variability of the wind direction, which would require continuous adjustments of the boundary conditions if the drone remains completely static, potentially causing an inlet boundary to behave like an outlet over time and rendering the simulation unstable. This issue can be easily addressed by setting a constant wind inlet direction and rotating the drone accordingly, which produces an analogous outcome. The rotation of the drone can be achieved using the Overset Mesh technique \cite{starccm_overset}, which allows for the drone rotation without modifying the direction of the inlet boundary conditions.

A drone traveling at a velocity $\mathbf{v}_{drone}$ through a wind field $\mathbf{v}_{wind}(\mathbf{r})$ is dynamically equivalent to a static drone subjected to a wind field equal to $\mathbf{v}_{wind}(\mathbf{r}) - \mathbf{v}_{drone}$, which is the principle that underlies the operation in wind tunnels. From this concept, we can extract the wind field along any given trajectory from the stationary urban simulation and introduce it into the wind tunnel, with a varying wind speed at the inlet. The drone rotates in real time to align with the wind direction, as experienced by the moving coordinate system traveling along the trajectory at velocity $\mathbf{v}_{drone}$.

This setup allows us to compare the results from the wind tunnel simulation to those obtained using the more time-consuming method of embedding the drone within the entire urban domain. Here, we detail the configuration of both simulations and present a comparative analysis. If a strong correlation is observed between the results, the proposed wind tunnel approach can serve as an efficient and faster alternative for evaluating wind loads and aerodynamic forces over optimized trajectories.

To demonstrate the flexibility of our urban reconstruction methodology, we employed a different scenario for the drone simulations compared to the one previously presented. For testing purposes, we selected a flat area with long, uninterrupted streets in the Cuatro Caminos district of Madrid, which allows for a straight, turn-free trajectory of sufficient length. This setup provided an ideal test-bed for validating our approach. Once validated, this methodology can be readily applied to virtually any drone path within an urban environment.

The drone simulation across the full urban domain is initialized using results from a previously conducted steady-state CFD simulation, which provides the baseline wind field. Implementing an Overset Mesh requires careful mesh matching between the background and overset regions to ensure consistent grid resolution, thereby increasing both the computational cost and setup complexity. An overview of the mesh setup is provided in Figure \ref{fig:mesh_wind_tunnel}(a).

For the wind tunnel simulation, a domain with the dimensions and mesh configuration shown in Figure \ref{fig:mesh_wind_tunnel}(b) is constructed. The drone geometry, adapted from a generic commercial model and rescaled to match the real size of the DJI Phantom 3 \cite{PAZ2020104378, PAZ2021104776}, is embedded using the same Overset Mesh technique, with its orientation dynamically updated at each time step to account for changing wind directions. Varying inlet velocities are imposed to emulate the drone's motion through the original urban wind field.

To ensure consistency between the full-domain and wind tunnel simulations, as illustrated in Figures \ref{fig:mesh_wind_tunnel}(c-d), a mesh independence study was conducted. We first identified an optimal mesh resolution within the spherical region around the drone in the wind tunnel setup, which was then carried over to the urban reconstruction. Three mesh sizes were tested, with detailed analysis of key reference metrics provided in the Supplementary Material. Since the drone is enclosed within a spherical Overset Mesh region in both simulations, a consistent mesh size was used throughout. Ensuring comparable donor and acceptor cell sizes necessitates pre-refinement of the drone's path region in the urban domain, significantly increasing cell count and computational load. The background mesh for the broader urban domain is consistent with that used previously for the calibration section, where a dedicated mesh independence assessment was also performed.

\section{Results}
\label{section:results}

First, we analyze the data extracted from multiple simulations using various meteorological prediction services, comparing the outcomes to real-world observations. Once the accuracy of the methodology is established, we proceed to compare the results from both drone simulations, showing the results compatibility and the difference in computational load.

\subsection{Calibration with meteorological station data}
\label{section:calibration}

\begin{figure}[t!]
	\centering
	\includegraphics[width=0.49\textwidth]{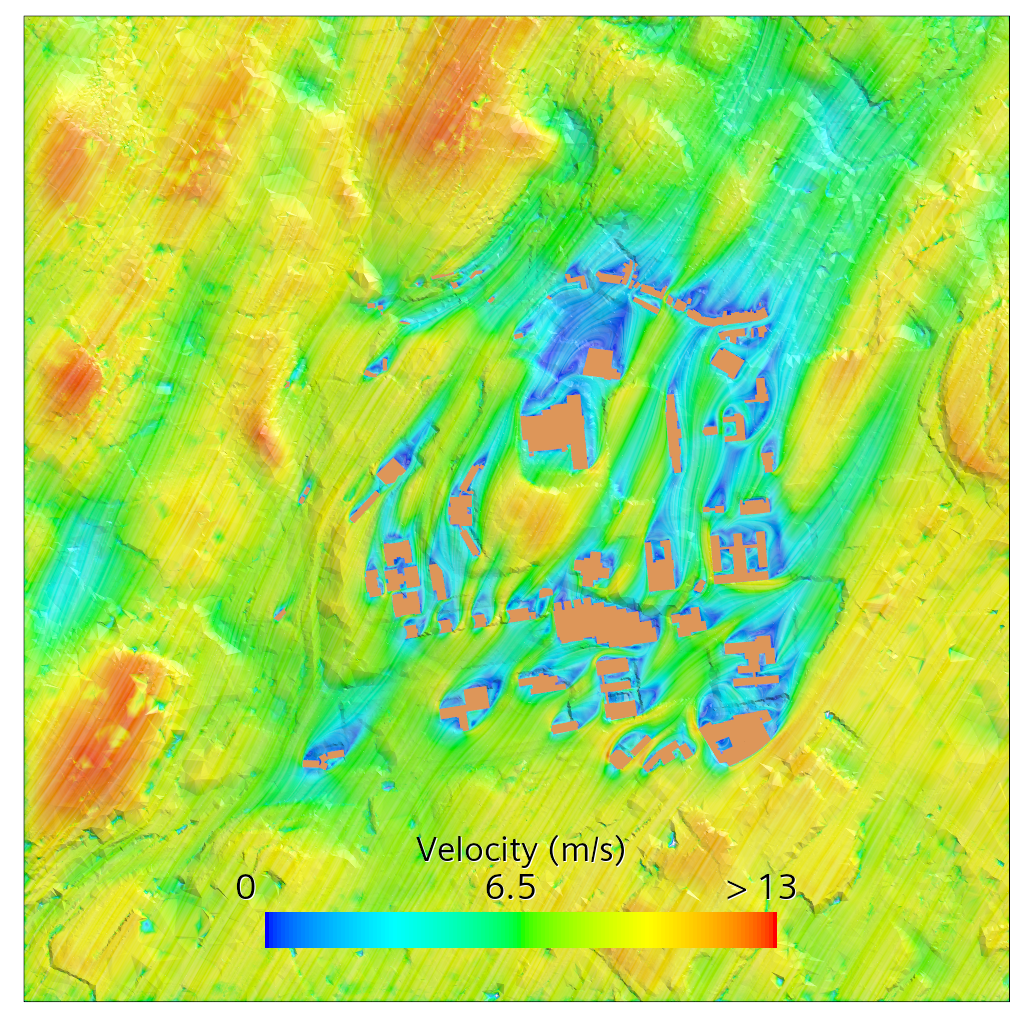}
	\includegraphics[width=0.49\textwidth]{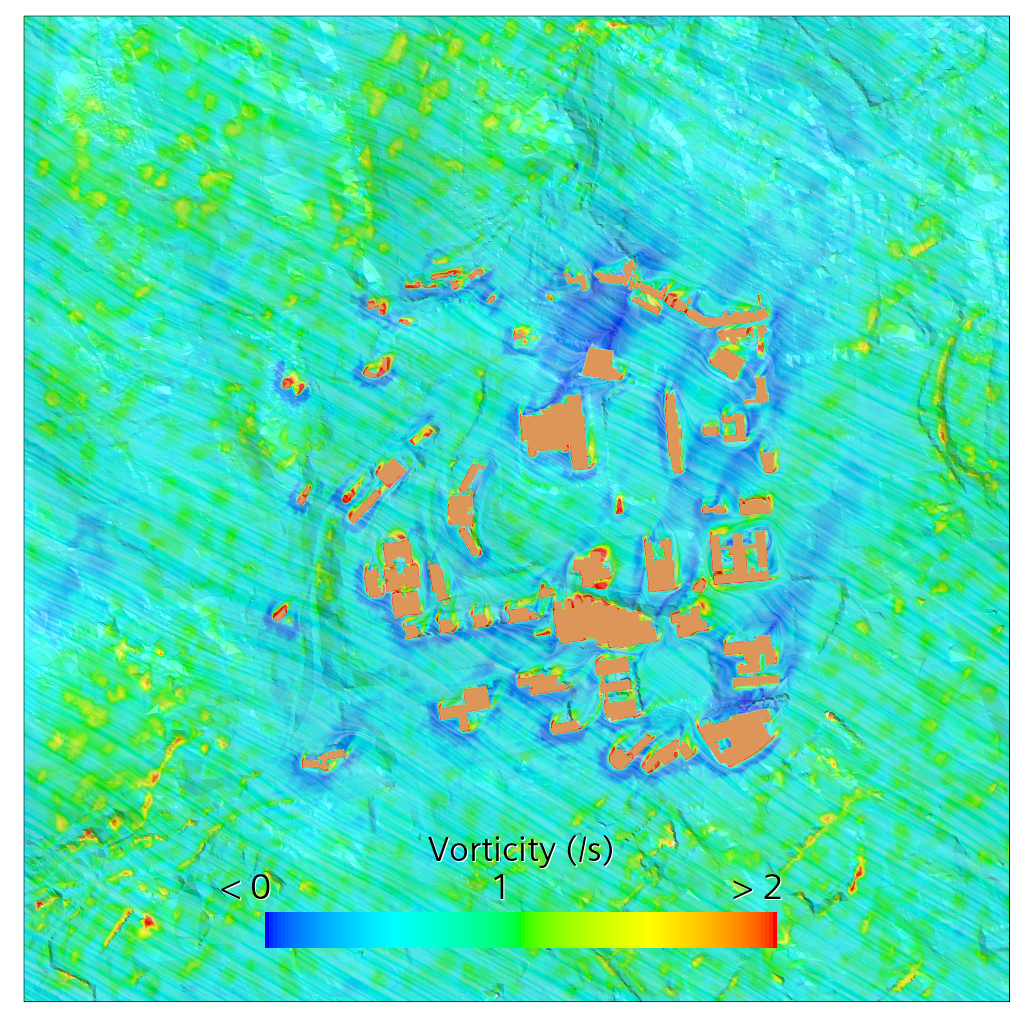}
	\caption{Visualization of the velocity and vorticity fields in a plane 5 m above ground level for an example simulation in our calibration domain at the Campus Sur of Santiago de Compostela, Spain.}
	\label{fig:velocity_plane}
\end{figure}

The meteorological predictions from the three weather forecast services used in our simulations are provided at different time scales. MeteoGalicia (Weather Research and Forecasting, WRF 1km model) and OpenMeteo provide wind speed and direction at a height of 10 meters every hour for a specific location, whereas AEMET predictions are provided in six-hour intervals with lower resolutions (5 km/h – 45º). While MeteoGalicia and AEMET predictions are uploaded at the start of each day and remain unchanged, OpenMeteo updates its predictions periodically. Consequently, downloading the predictions from OpenMeteo for 9:00 AM after that hour has passed will yield corrected predictions, which enables us to use OpenMeteo data as low-resolution validation data.

\begin{figure}[b!]
	\centering
	\includegraphics[width=1.0\textwidth]{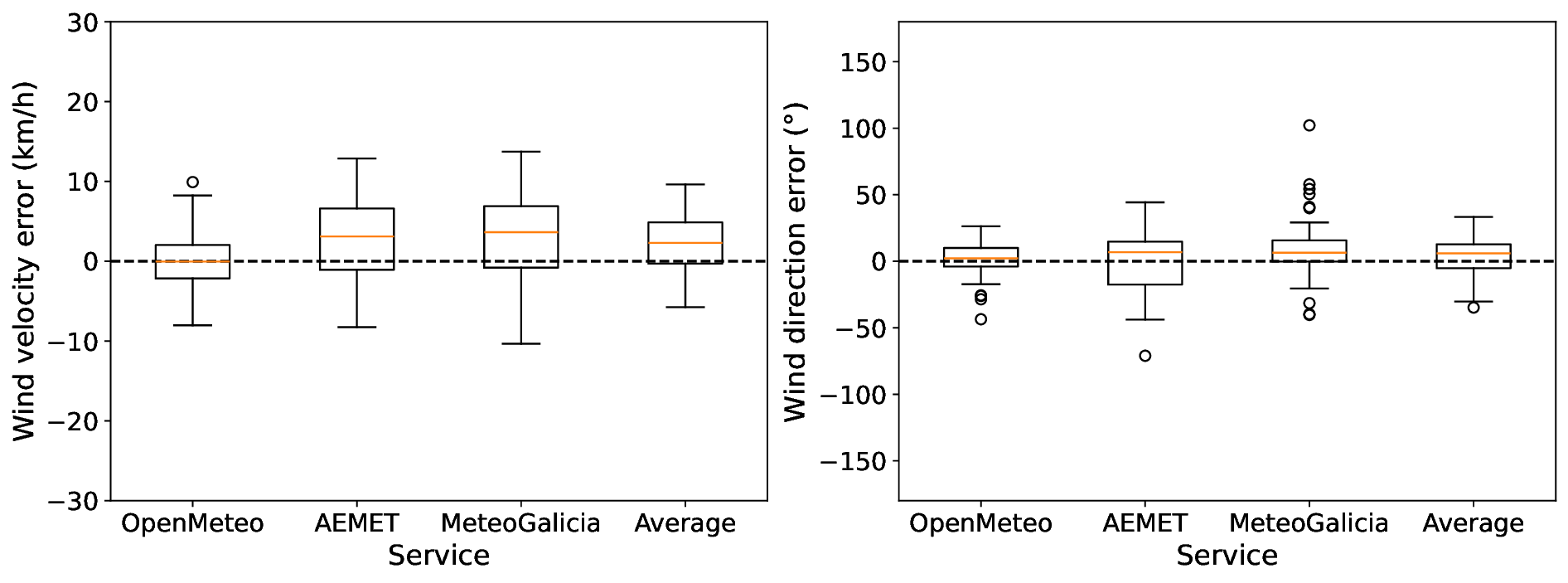}
	\caption{Boxplot of wind speed and direction errors at 10 m for each model considering the averaged real-life values over an hour around 9:00 AM.}
	\label{fig:boxplots}
\end{figure}

Incorporating these corrected predictions into our model should give closer results to the real data obtained from the meteorological station, while the other two services may provide less accurate results as they are based on future predictions. If the corrected OpenMeteo prediction agrees with the observation data, this means our model is working as expected.

The meteorological station employed as a reference for validating our simulations is positioned in the center of our wind tunnel (see Figure \ref{fig:geometry}). This station provides valuable information on wind speed and its direction (along with their corresponding standard deviations) every 10 minutes. Therefore, for every prediction from the MeteoGalicia and OpenMeteo models, the station records up to six measurements of the wind speed. To work with similar time scales in both prediction and observation, and due to the different time discretizations of our predictions, we will use the mean values from 8:30 to 9:30 AM as our real-life measurements, with their corresponding uncertainties.

\begin{figure}[t!]
	\centering
	\includegraphics[width=1.0\textwidth]{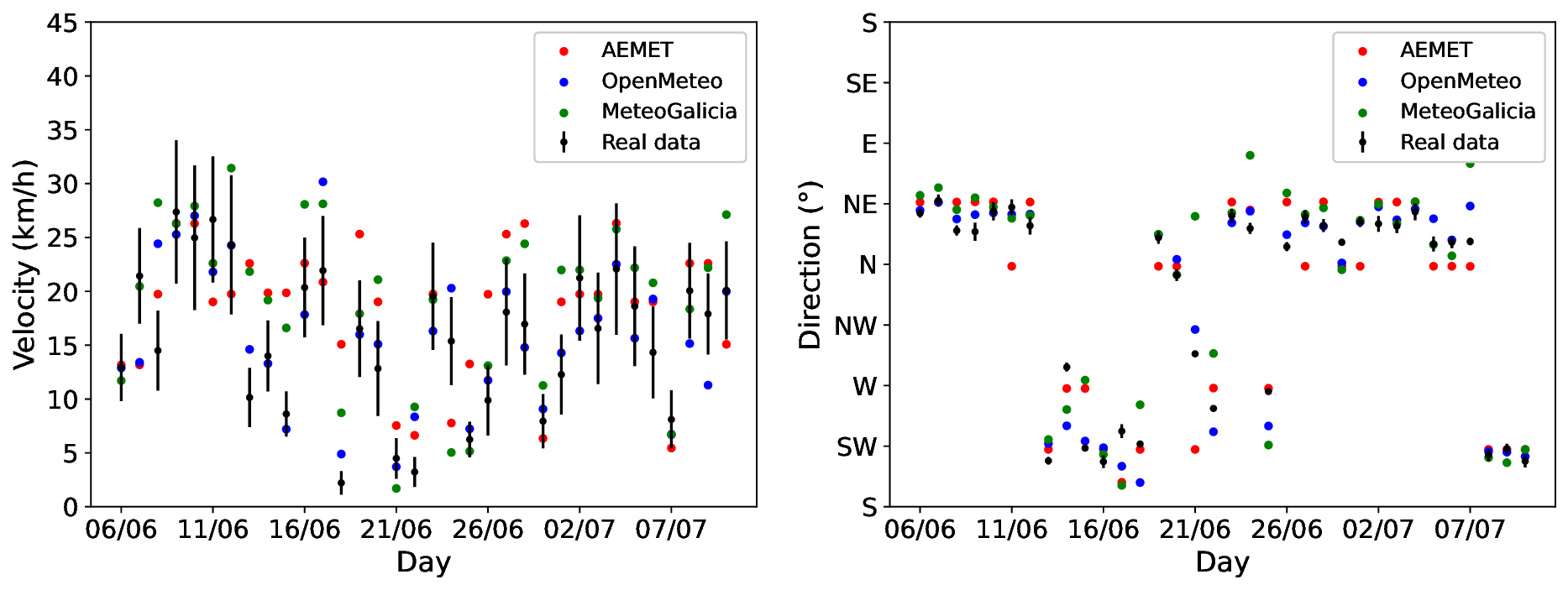}
	\caption{Comparison of each daily simulation (at 9:00 AM) with its corresponding real-life measurements.}
	\label{fig:data}
\end{figure}

Inside the simulations, a reference point was positioned at a height of ten meters above ground level on the exact coordinate of the meteorological station. This allowed us to extract wind speed and direction data at that specific point for a direct comparison with the real-life measurements. We can see the velocity and vorticity fields for an example simulation in Figure \ref{fig:velocity_plane}. The average errors for each prediction model are displayed in boxplots in Figure \ref{fig:boxplots}, and a complete comparison for each individual real-life measurement and simulation values is presented in Figure \ref{fig:data}. As evidenced by the boxplots in Figure \ref{fig:boxplots}, OpenMeteo predictions are consistently more accurate than those from the other two services. This was expected, given that OpenMeteo predictions are corrected throughout the day, thereby reproducing sudden changes in meteorological conditions.

\begin{figure}[t!]
	\centering
	\includegraphics[width=1.0\textwidth]{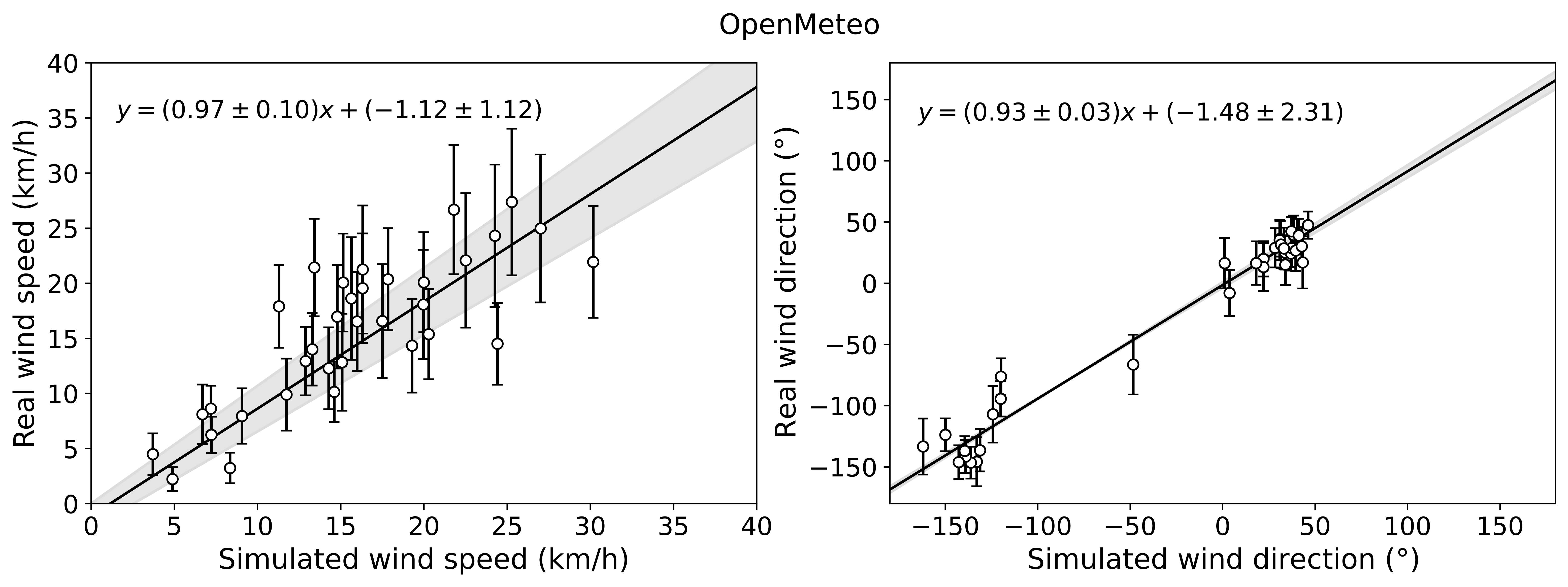}
	\includegraphics[width=1.0\textwidth]{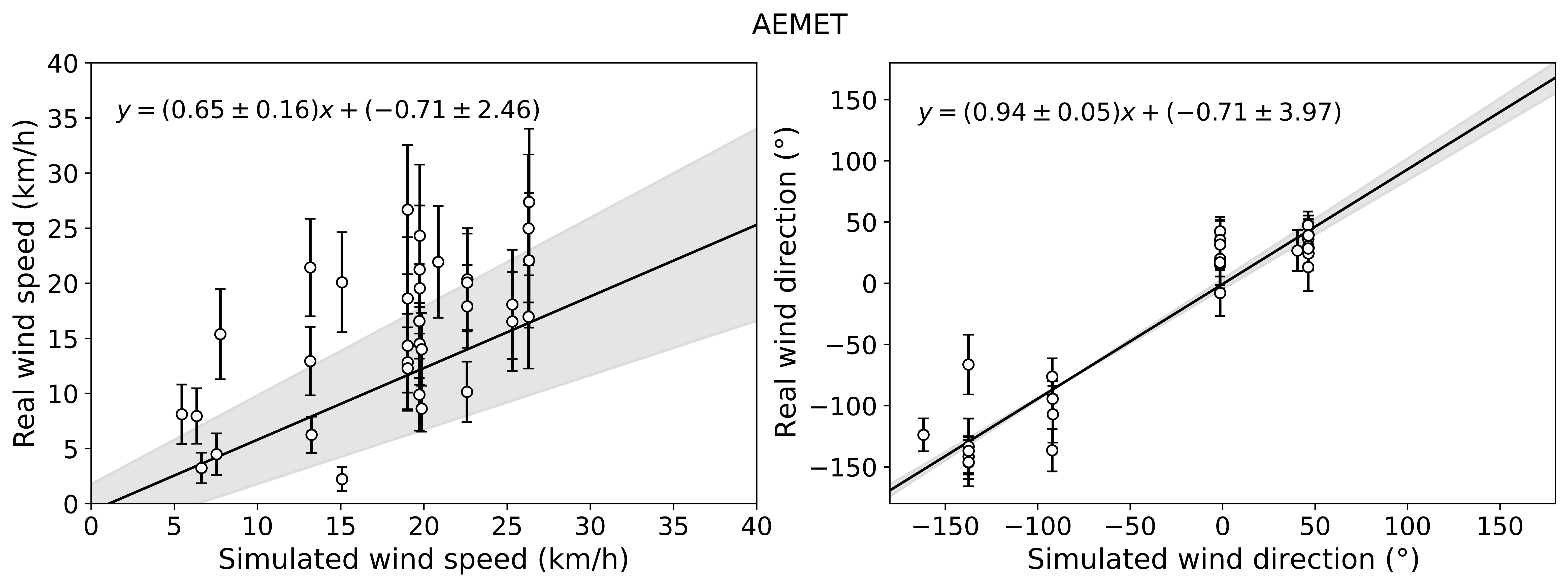}
	\includegraphics[width=1.0\textwidth]{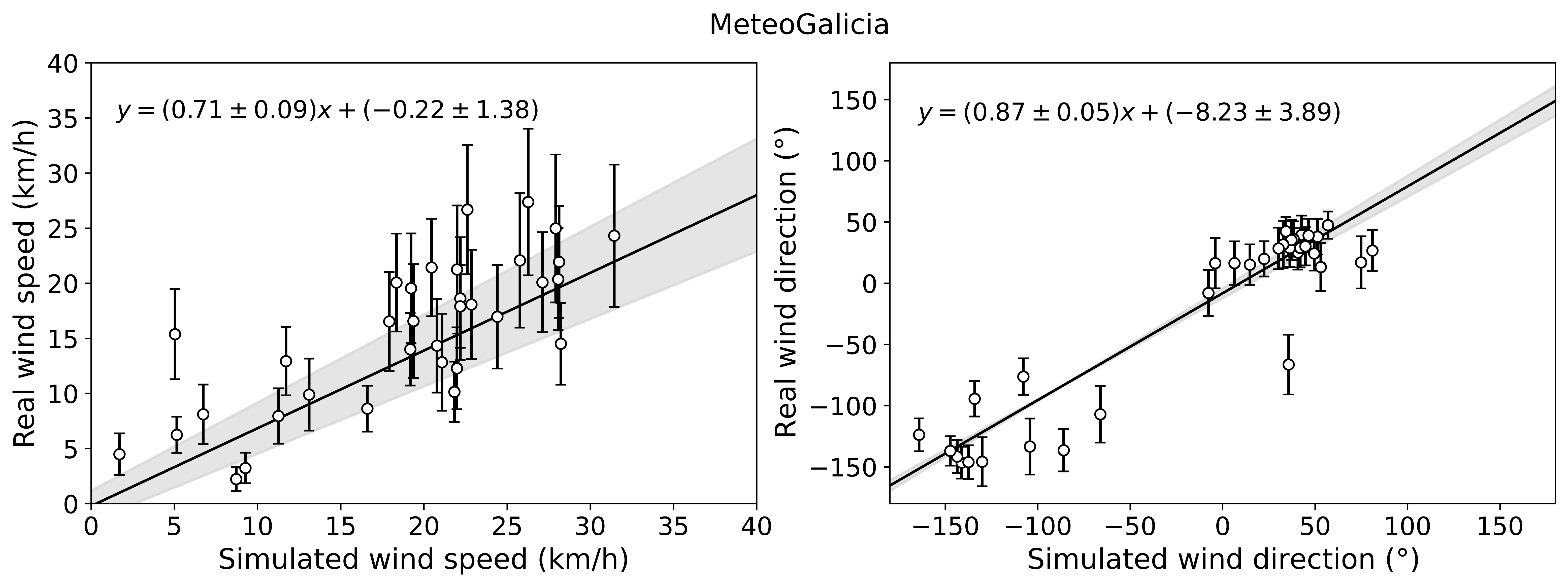}
	\caption{Linear regression of the simulated and measured wind speeds (left column) and wind directions (right column) for the three weather services (OpenMeteo, AEMET, and MeteoGalicia). The gray area represents the highest and lowest possible straight lines using the parameter uncertainties of the linear fit, calculated as the square root of the diagonal components of the covariance matrix.}
	\label{fig:linear_fits}
\end{figure}

A simple way to compare the meteorological station measurements with the simulation results is to plot them together for each prediction service. If the simulations are correct, the points should follow a straight line of slope equal to one and an independent term equal to zero. A linear fit, accounting for the given uncertainties, was performed for each case and is presented in Figure \ref{fig:linear_fits}.

The linear fit for the OpenMeteo data is best adjusted to a slope close to one, with an independent term close to zero, both for wind speed and direction. This means that our model for the geometry generation and the creation of boundary conditions for the inlet is working correctly. Consequently, when low resolution data from meteorological predictions is provided, our model can yield accurate local results in terms of the simulated wind speed and direction.

The MeteoGalicia and AEMET predictions are less accurate, as they are not updated in subsequent hours, leaving us with past predictions that have higher error rates. Nevertheless, we observe good predictions for wind direction, whereas wind speed is more challenging to predict due to wind gusts that can affect the results. Despite this, there is a satisfactory agreement for wind speeds in both services, indicated by the positive slopes. The results may be influenced by the lower uncertainties associated with lower wind speed values, which could play a significant role in the linear fits. Both predictions tend to overestimate wind speed, as evidenced by slopes lower than one. This overestimation is preferable since it means we are considering the worst-case scenarios. The spatial and temporal resolution of MeteoGalicia predictions (1 hour) is finer than that of AEMET predictions (6 hours), so the superior results from MeteoGalicia were anticipated.

There are some days in which all three prediction models fail systematically, either in terms of wind speed or direction. However, in other occasions, only OpenMeteo accurately reflects the correct wind components. For instance, on June 21st (see Figure \ref{fig:data}), we could see that the wind direction was very unstable. Examining the registered values from 00:00 AM to 7:00 AM, we observe momentary fluctuations in speed, oscillating around different values. At approximately 9:00 AM, there is a sudden change in direction, and only OpenMeteo successfully reproduces the correct value. In contrast, the other two services indicate completely opposite directions. This discrepancy is largely due to the differing timescales of the predictions; one service predicts the change will occur later, while the other forecasts it earlier.

In the preceding days from July 5th and 6th, the wind consistently registered northeast (NE) component. In those two days, the wind direction shifted slightly closer to the north (N), falling between NE and N. This shift resulted in a change in the AEMET prediction, which adjusted from forecasting a direction of 45º (NE) to 0º (N) when this minor change occurred. This adjustment demonstrates the capability of AEMET, as it accurately predicted this small change in wind direction despite its lower resolution in direction and speed. Although AEMET's forecasts are not always precise due to this resolution limitation, it is noteworthy that such a subtle shift was captured even by AEMET, which generally provides the least accurate wind speed predictions.

To quantitatively evaluate the agreement between the simulation results and the real-world measurements, beyond basic linear regression, we employ the Concordance Correlation Coefficient (CCC). Unlike the Pearson correlation coefficient, which only measures the strength of the linear association, the CCC assesses both accuracy (closeness to the identity line) and precision (scatter around the best-fit line). It is therefore ideal for our case, where the goal is to reproduce real-world values as closely as possible using simulated data. The CCC penalizes deviations from both perfect correlation and the ideal 1:1 relationship, making it a comprehensive measure of agreement between predicted and observed wind conditions. The coefficient is defined as,

\begin{equation}
	\label{eq:ccc}
	\rho_c = \frac{2\rho\sigma_x \sigma_y}{\sigma_x^2 + \sigma_y^2 + \left(\mu_x - \mu_y\right)^2}
\end{equation}
where $\rho$ is the Pearson correlation coefficient, $\mu_x$ and $\mu_y$ are the means of the predicted and observated values, respectively, and $\sigma_x$ and $\sigma_y$ are their standard deviations. A CCC value of 1 indicates perfect agreement, while values closer to 0 indicate poorer concordance.

\begin{table}[h!]
	\centering
	\caption{CCC values for each prediction service.}
	\label{tab:CCC}
	\begin{tabular}{c|c|c|}
		\cline{2-3}
		& Wind speed & Wind direction \\ \hline
		\multicolumn{1}{|c|}{OpenMeteo} & 0.853 & 0.985 \\ \hline
		\multicolumn{1}{|c|}{AEMET} & 0.627 & 0.959 \\ \hline
		\multicolumn{1}{|c|}{MeteoGalicia} & 0.726 & 0.934 \\ \hline
	\end{tabular}%
\end{table}

The extracted CCC values of wind speed and direction for each prediction service can be seen in Table \ref{tab:CCC}. Consistently with the previous discussion, we can see a very good agreement for the OpenMeteo service, while MeteoGalicia and AEMET behave worse given the future-like predictions.

\subsection{Wind tunnel for a moving drone inside the city}
\label{section:wind_tunnel}

In section \ref{section:meth_wind_tunnel}, we explained the problematic when trying to find the behavior of a moving drone embedding it inside the whole reconstructed domain. The mesh used for the stationary simulations had to be refined all along the drone trajectory in a now transient simulation with constant boundary conditions.

To overcome this limitation, we proposed extracting the values from a fast stationary simulation of the city and then using them inside a wind tunnel with a rotating drone and varying inlet velocities replicating the wind currents that the drone encounters while traveling.

\begin{figure}[t!]
	\centering
	\includegraphics[width=0.7\textwidth]{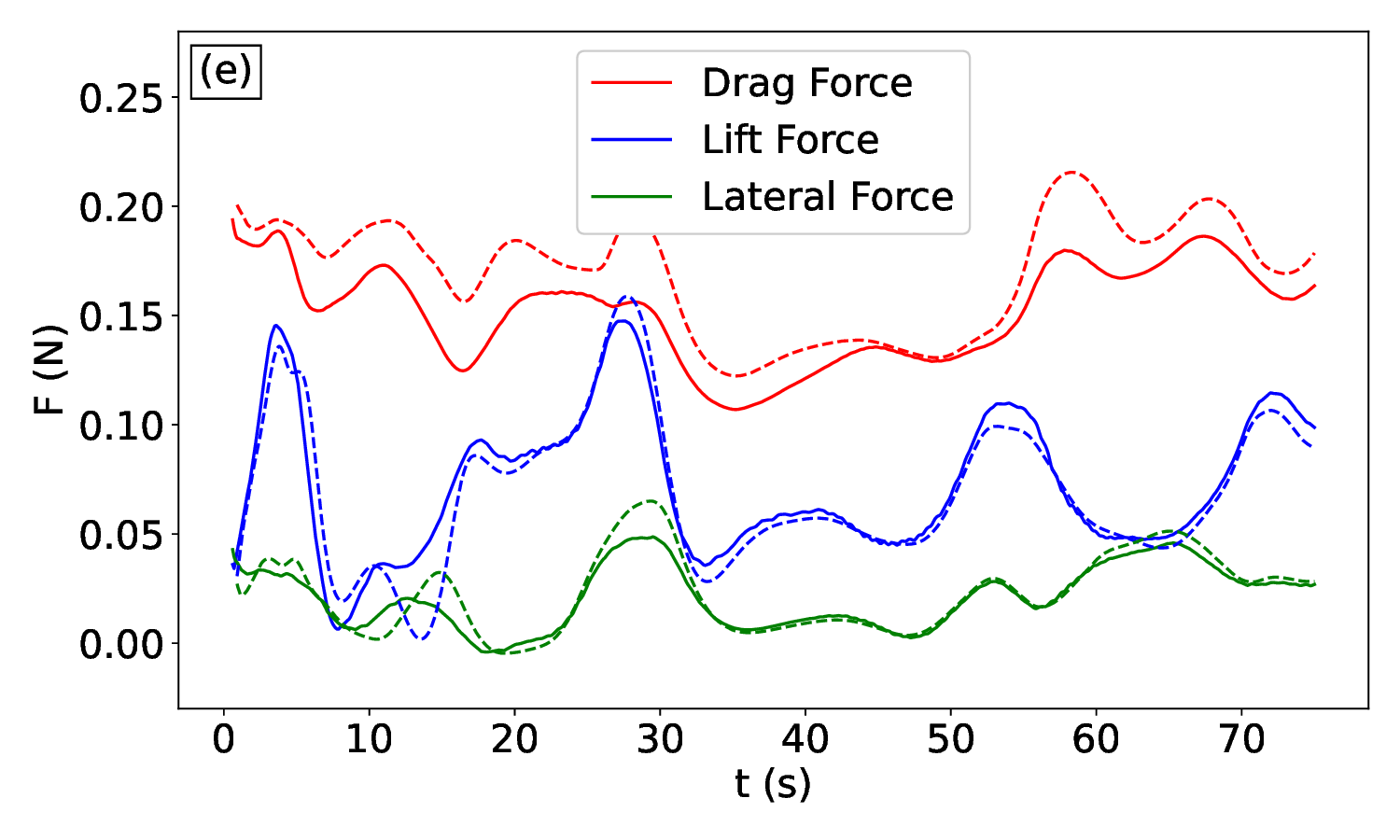}
	\caption{Comparison of the forces acting over the drone geometry using both approaches. The dotted lines represent the results from the city reconstruction, while the full lines represent the wind tunnel test.}
	\label{fig:results_wt}
\end{figure}

The computational time gain is, by far, the greatest reason to use this approach. The execution of the stationary city reconstruction plus the transient wind tunnel simulation was done in less than two hours (around 20 minutes for the city reconstruction and one hour and a half for the wind tunnel), while all the execution and configuration of the other approach took more than a day. Keep in mind that performing the simulation in a different path or scenario would not yield faster results, as we would have to reconfigure the mesh and all the drone settings, while the wind tunnel is already configured except for the boundary conditions, which are immediately extracted using a Python program.

However, this computational gain would be useless if the results of both simulations were not comparable. In Figure \ref{fig:results_wt} we can see a comparison between both methodologies.

As we can see, results from both simulations are nearly the same, mainly in the lateral and lift forces, where the difference between both methods is certainly small. The drag force seems to be consistently greater for the whole city simulation, in which some of the oscillations during flight are intensified. This can be caused due to the intrinsic instability of transient simulations, in which the size of the mesh might cause some variations even with constant boundary conditions. We have to keep in mind the great difference in mesh resolution between both cases, as the extracted wind data for the wind tunnel has a coarser mesh. Although mesh independency tests were performed, slight differences are to be expected. Other factors, such as the number of inner iterations or time step can explain the slight differences between both methods.

Overall, we observe a good agreement between the methodologies, which allowed us to verify the correct behavior of the wind tunnel test, reducing the simulation times of the drone inside the city domain, while maintaining very accurate results. Thus, the wind tunnel test can be used as a fast tool to evaluate the loads of the drone following any complex trajectory within a reconstructed city domain, immensely simplifying the simulation setup process.

\section{Summary and conclusions}

This study aimed to assess the accuracy of different meteorological prediction models in reconstructed urban environments simulated via CFD, by comparing them with real-life measurements. The method included not only the man-made constructions but also the nature of the surrounding plants. We successfully integrated the reconstruction of CFD urban simulations with different meteorological predictions, providing an analysis on the airflow dynamics. The integration of high-resolution urban geometries and meteorological data improves the accuracy of CFD simulations, making them reliable for practical applications. Our analysis showed that corrected predictions have a higher level of accuracy, showing that the proposed model is, indeed, a great tool to predict wind speeds in urban environments.

Our analysis revealed that corrected meteorological predictions yield greater accuracy, confirming the effectiveness of the proposed model in forecasting wind speeds in complex urban settings. A linear fit analysis indicated a strong correlation between OpenMeteo data and ground truth measurements, highlighting the robustness of our geometry reconstruction and boundary condition generation methods. The model performed well even when provided with low-resolution input data, successfully predicting local wind speed and direction.

Conversely, future forecasts from MeteoGalicia and AEMET---both of which are not updated beyond their initial 0:00 AM prediction---exhibited higher error margins. Nonetheless, these models still provided reasonable estimations of wind direction and speed, although they tended to slightly overestimate wind intensity. Interestingly, this overestimation can be advantageous in urban airflow simulations by ensuring conservative, worst-case scenario modeling, which is often desirable in safety-critical applications.

The proposed methodology has broad applicability across fields such as urban planning, environmental monitoring, and unmanned aerial vehicle (UAV) operations. Validation against ground-truth meteorological station data supports the credibility of the simulations and enhances confidence in their deployment for real-world decision-making. With the increasing global availability of LiDAR data, this approach can be easily adapted to cities worldwide, contributing to more informed urban management and planning strategies.

We also applied this automated methodology to a moving drone scenario to simulate aerodynamic forces acting on the drone as it traversed the generated velocity field. We compared two modeling strategies: embedding the drone within the urban CFD simulation and using a wind tunnel-style test. The wind tunnel approach delivered comparable results but with significantly faster computational times, underscoring the practical benefits of our framework.

In summary, this research presents a robust, adaptable methodology for integrating CFD simulations with meteorological predictions in urban environments. It proves effective in accurately reconstructing airflow patterns and offers practical utility for various real-world applications. The ability to quickly generate reliable wind fields and simulate their impact on dynamic elements like UAVs marks a significant advancement in urban environmental modeling.

\section*{Authors' contributions}
MSV designed the model, programmed the code, performed the CFD simulations, analyzed the results and wrote the original draft. All other authors (SVB, AOC, APM and JM) participated equally on supervision, conceptualization, and review of the manuscript.

\section*{Declaration of competing interest}
The authors declare that they have no known competing financial interests or personal relationships that could have appeared to influence the work reported in this document.

\section*{Acknowledgments}
Some simulations were run using the Supercomputer Center of Galicia (CESGA) and we acknowledge their support. M. Suárez-Vázquez thanks the support of the Doutoramento Industrial program from GAIN-Xunta de Galicia (IN606D).

\bibliographystyle{unsrt}
\bibliography{sample}

\end{document}


\date{}
	\maketitle
	
	\keywords{Computational fluid dynamics, urban planning, terrain reconstruction, building
		reconstruction, validation study.}

	\section{Reconstruction of other regions}
	\label{reconstruction}
	
	To show the generalizability of the developed methodology, we attach a screenshot of some reconstructions performed for other Spanish locations.
	
	\begin{figure}[h!]
		\centering
		\includegraphics[width=0.95\textwidth]{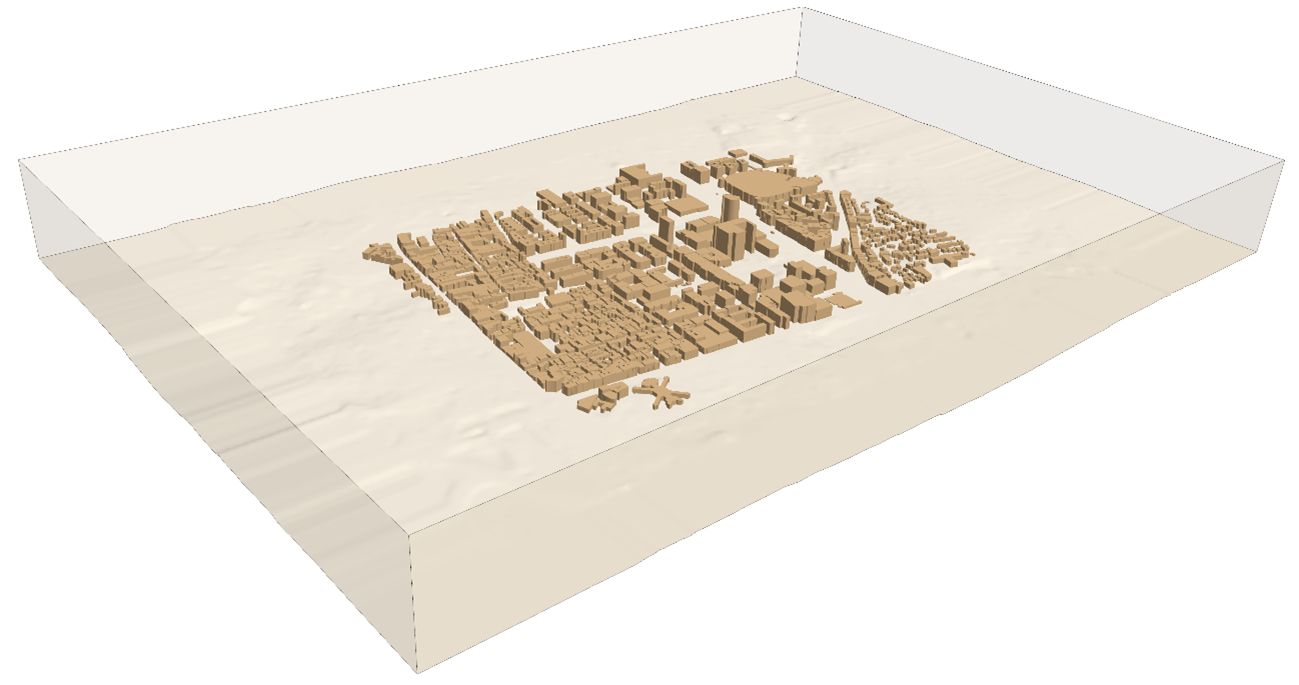}
		\caption{Reconstruction of the Chamber\'i neighborhood in Madrid.}
		\label{fig:madrid}
	\end{figure}
	
	\begin{figure}[h!]
		\centering
		\includegraphics[width=0.95\textwidth]{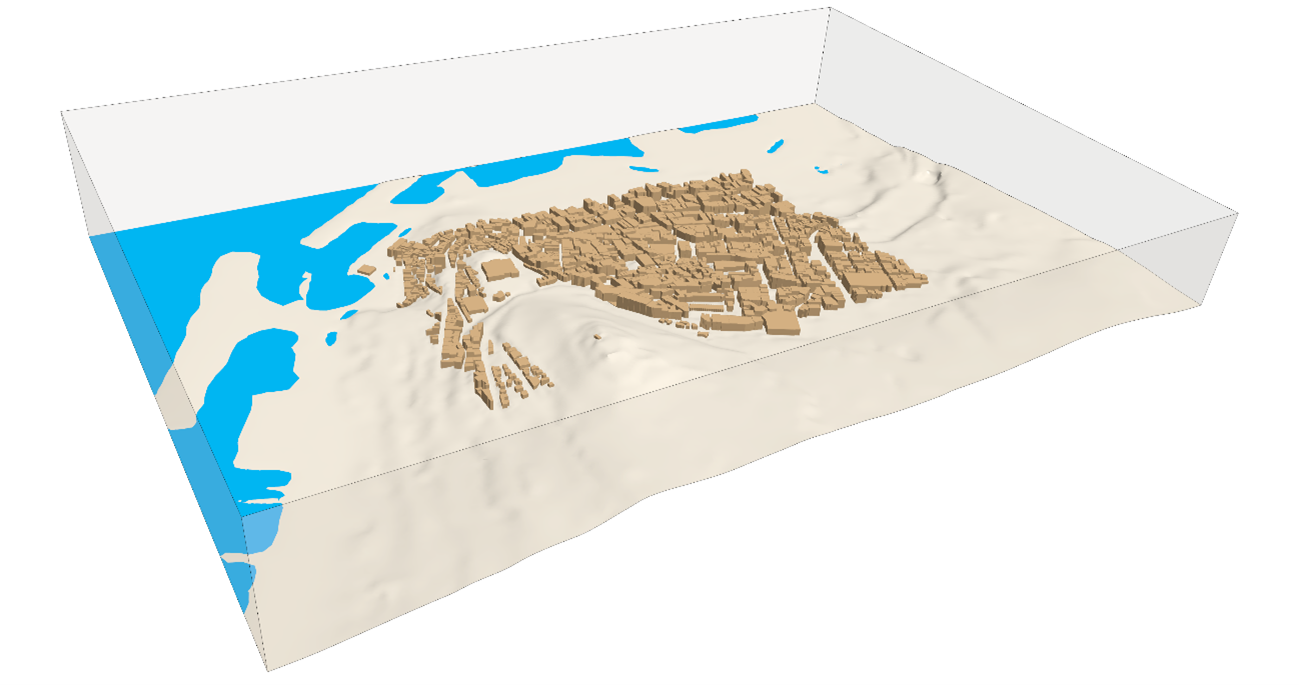}
		\caption{Reconstruction of the old part of Vigo, near the port. The blue part represents the sea.}
		\label{fig:vigo}
	\end{figure}
	
	\begin{figure}[h!]
		\centering
		\includegraphics[width=0.95\textwidth]{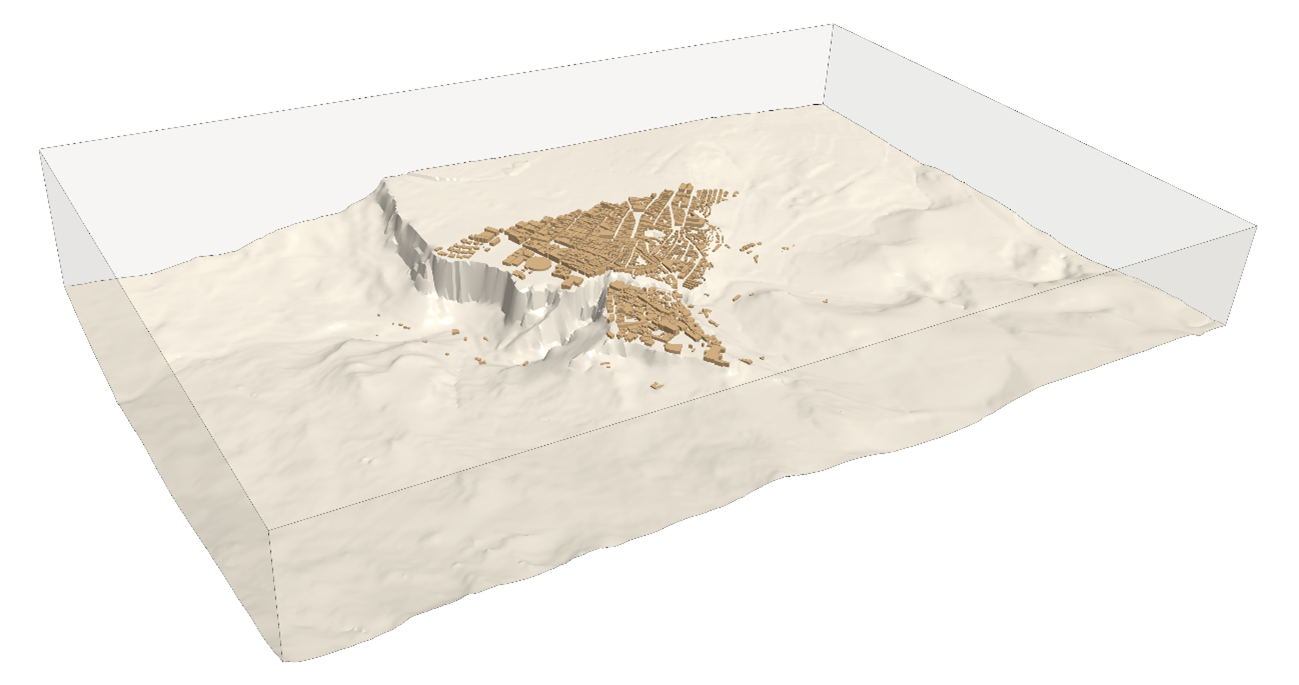}
		\caption{Reconstruction of the city of Ronda, in Málaga. This city is known for being built over a high cliff, which is correctly reconstructed by our method.}
		\label{fig:ronda}
	\end{figure}
	
	\section{Mesh independency tests}
	
	Here we provide the extracted data for both sets of simulations performed in our analysis.
	
	\subsection{Urban reconstrucion simulations}
	
	In Table \ref{tab:mesh}, a summary of the important magnitudes used for the mesh independency test can be found. We extracted the wind velocity and direction in two points (one situated at the meteorological station position and the other near the area with most buildings). We also extracted other quantities in different surfaces and area, but we do not show them here for brevity.
	
	\begin{table}[h!]
		\centering
		\caption{Relative error and cell count for different base sizes}
		\vspace{1mm}
		\begin{tabular}{|c|c|c|c|}
			\hline
			\textbf{Base size (m)} & \textbf{Number of cells} & 
			\makecell{\textbf{Relative error} \\ \textbf{(\%) - Point 1}} & 
			\makecell{\textbf{Relative error} \\ \textbf{(\%) - Point 2}} \\
			\hline
			8.0  & 3.96E+06 & -0.29 & 0.60 \\
			10.0 & 2.86E+06 & -0.24 & -0.01 \\
			12.0 & 2.38E+06 & -1.19 & 0.49 \\
			16.0 & 1.63E+06 & ---   & ---  \\
			20.0 & 1.36E+06 & 2.88  & 1.00 \\
			\hline
		\end{tabular}
		\label{tab:mesh}
	\end{table}
	
	As we can see, a Base Size of 16.0 m can be taken as a good compromise between the number of cells and the errors in the simulations. We see that all the reference values have a relative difference lower than 1\%, even for the finest mesh. The coarsest mesh (Base Size, 20.0 m) raises the relative error to a 3\%, so we will use the next one (Base Size, 16.0 m) as it has nearly the same number of cells but with lower relative differences with respect to finer meshes.
	
	\subsection{Wind tunnel simulations}
	
	For the case of the wind tunnel simulations, a mesh independency test of the spherical domain surrounding the drone was performed. The background mesh of the overset process was chosen so that the size of the cells was approximately equal to the one for the outer cells of the spherical overset domain. It is important to notice that this sphere has the same size than the one used inside the big transient simulation, so this test can be also applicable to it without needing to perform another one. For the background mesh, we can use the previous urban reconstruction mesh independency test. Three different mesh sizes for this case were considered, and results can be seen in Table \ref{mesh2}.
	
	\begin{table}[h!]
		\centering
		\caption{Forces and pressure values for different base sizes.}
		\resizebox{\textwidth}{!}{%
		\vspace{1mm}
		\begin{tabular}{|c|c|c|c|c|}
			\hline
			\textbf{Base size (m)} & \textbf{Drag force (N)} & \textbf{Lateral force (N)} & \textbf{Lift force (N)} & \textbf{Pressure (Pa)} \\
			\hline
			0.1 & 0.20 & 0.0059  & 0.072 & -3.89 \\
			0.2 & 0.20 & 0.0064  & 0.071 & -3.84 \\
			0.3 & 0.21 & 0.0089  & 0.076 & -3.55 \\
			\hline
		\end{tabular}
		}
		\label{mesh2}
	\end{table}
	
	The results show that the 0.1 m and 0.2 m Base Sizes offer very similar values in all the magnitudes analyzed, while the 0.3 m presents more notable deviations, especially in lateral force and pressure. Therefore, the 0.2 m base size is chosen, since it maintains good accuracy with a considerably smaller number of cells than the 0.1 m.